\documentclass[twocolumn]{aastex631}
\usepackage{graphicx}   
\usepackage{lineno}
\usepackage{bm}

\shorttitle{Cosmology with one galaxy?-II}
\shortauthors{Echeverri, Villaescusa-Navarro et al.}
\graphicspath{{./}{Graficas/}}
\begin{document}

\title{Cosmology with one galaxy? - The ASTRID model and robustness}

\author{Nicolas Echeverri}
\affiliation{Instituto de Fisica, Universidad de Antioquia, A.A.1226, Medellin, Colombia}

\author[0000-0002-4816-0455]{Francisco Villaescusa-Navarro}
\affiliation{Center for Computational Astrophysics, Flatiron Institute, 162 5th Avenue, New York, NY, 10010, USA}
\affiliation{Department of Astrophysical Sciences, Princeton University, Peyton Hall, Princeton NJ 08544, USA}

\author{Chaitanya Chawak}
\affiliation{Indian Institute of Science Education and Research (IISER) Tirupati, Tirupati-517507, India}

\author[0000-0001-7899-7195]{Yueying Ni}
\affiliation{Harvard-Smithsonian Center for Astrophysics, 60 Garden Street, Cambridge, MA 02138, US}
\affiliation{McWilliams Center for Cosmology, Department of Physics, Carnegie Mellon University, Pittsburgh, PA 15213, US}

\author[0000-0003-1197-0902]{ChangHoon Hahn}
\affiliation{Department of Astrophysical Sciences, Princeton University, Peyton Hall, Princeton NJ 08544, USA}

\author[0000-0002-1329-9246]{Elena Hern\'andez-Mart\'inez}
\affiliation{Universit\"ats-Sternwarte, Fakult\"at f\"ur Physik, Ludwig-Maximilians-Universit\"at M\"unchen, Scheinerstr. 1, 81679 M\"unchen, Germany}

\author[0000-0001-7689-0933]{Romain Teyssier}
\affiliation{Department of Astrophysical Sciences, Princeton University, 4 Ivy Lane, Princeton, NJ 08544 USA}

\author[0000-0001-5769-4945]{Daniel Anglés-Alcázar}
\affiliation{Center for Computational Astrophysics, Flatiron Institute, 162 5th Avenue, 
New York, NY, 10010, USA}
\affiliation{Department of Physics, University of Connecticut, 196 Auditorium Road, U-3046, 
Storrs, CT, 06269, USA}

\author{Klaus Dolag}
\affiliation{Universit\"ats-Sternwarte, Fakult\"at f\"ur Physik, Ludwig-Maximilians-Universit\"at M\"unchen, Scheinerstr. 1, 81679 M\"unchen, Germany}
\affiliation{Max-Planck-Institut f\"ur Astrophysik, Karl-Schwarzschild-Stra{\ss}e 1, 85741 Garching, Germany}

\author[0000-0002-6292-3228]{Tiago Castro}
\affiliation{INAF-Osservatorio Astronomico di Trieste, Via G. B. Tiepolo 11, I-34143 Trieste, Italy}
\affiliation{INFN, Sezione di Trieste, Via Valerio 2, I-34127 Trieste TS, Italy}
\affiliation{IFPU, Institute for Fundamental Physics of the Universe, via Beirut 2, 34151 Trieste, Italy}

\correspondingauthor{Nicolas Echeverri}
\email{nicolas.echeverrir@udea.edu.co}

\begin{abstract}

Recent work has pointed out the potential existence of a tight relation between the cosmological parameter $\Omega_{\rm m}$, at fixed $\Omega_{\rm b}$, and the properties of individual galaxies in state-of-the-art cosmological hydrodynamic simulations. In this paper, we investigate whether such a relation also holds for galaxies from simulations run with a different code that made use of a distinct subgrid physics: Astrid. We find that also in this case, neural networks are able to infer the value of $\Omega_{\rm m}$ with a $\sim10\%$ precision from the properties of individual galaxies while accounting for astrophysics uncertainties as modeled in CAMELS. This tight relationship is present at all considered redshifts, $z\leq3$, and the stellar mass, the stellar metallicity, and the maximum circular velocity are among the most important galaxy properties behind the relation. In order to use this method with real galaxies, one needs to quantify its robustness: the accuracy of the model when tested on galaxies generated by codes different from the one used for training. We quantify the robustness of the models by testing them on galaxies from four different codes: IllustrisTNG, SIMBA, Astrid, and Magneticum. We show that the models perform well on a large fraction of the galaxies, but fail dramatically on a small fraction of them. Removing these outliers significantly improves the accuracy of the models across simulation codes.
    
\end{abstract}

\keywords{Cosmological parameters --- Galaxy processes --- Computational methods --- Astronomy data analysis --- Neural Networks}

\section{Introduction} 
\label{sec:intro}

Inferring the value of the cosmological parameters is one of the most essential tasks in cosmology. Galaxy clustering is commonly used to carry out this task, although other methods (e.g. the cosmic distance ladder) can also be used to estimate the value of some parameters. 

Recently, \citet{Cosmo1gal} claimed that the properties of individual galaxies could be used to infer the value of $\Omega_{\rm m}$. The authors showed that by training neural networks on galaxy properties from individual galaxies to perform likelihood-free inference on the value of the cosmological parameters, they were able to constrain $\Omega_{\rm m}$, at fixed $\Omega_{\rm b}$, with $\sim10\%$ precision. Authors presented a potential explanation stating that galaxy properties live in a low-dimensional manifold that is affected differently by $\Omega_{\rm m}$ than by astrophysical processes such as supernovae and active galactic nuclei (AGN) feedback. In that work, the authors used thousands of state-of-the-art hydrodynamic simulations from the Cosmology and Astrophysics with MachinE Learning (CAMELS) project \citep{CAMELS, CAMELS_public}. These included simulations performed with the AREPO \citep{Arepo} and GIZMO \citep{Hopkins2015_Gizmo} hydrodynamic codes implementing the subgrid galaxy formation models of IllustrisTNG \citep{illustris:Nelson-2015} and SIMBA \citep{Dave2019_Simba}. 

In this work we made use of a new suite of simulations, CAMELS-Astrid \citep{CAMELS_Astrid}, run with the MP-Gadget code using the Astrid model \citep{Astrid1, Astrid2}, that solve the hydrodynamic equations and implement feedback with a yet different method than the CAMELS-IllustrisTNG and CAMELS-SIMBA simulations discussed above. As with CAMELS-IllustrisTNG and CAMELS-SIMBA, the CAMELS-Astrid simulations have different values of the cosmological and astrophysical parameters.

We show that neural networks can also infer the value of $\Omega_{\rm m}$, at fixed $\Omega_{\rm b}$, from the properties of individual galaxies of the CAMELS-Astrid simulations with a $\sim10\%$ precision. We also investigate what are the most important galaxy properties used by the model to make the inference and show that our results hold at different redshifts. We then focus our attention on the robustness of the different models (i.e. models trained on galaxies from different simulations). We show that the models perform well on most galaxies, and removing outliers help the model to make unbiased predictions.

This paper is organized as follows. In Sec. \ref{sec:methods} we describe the data and machine learning methods we use. Next, we present the results of our analysis, in terms of precision and accuracy of the models, in Sec. \ref{sec:results}. We then conclude in Sec. \ref{sec:summary}.

\section{Methods} 
\label{sec:methods}

In this section we describe the data we use and the machine learning methods we utilize. We also outline the metrics we consider to quantify the accuracy and precision of the models.

\subsection{Data}

We train our model using galaxy properties from individual galaxies of the CAMELS hydrodynamic simulations \citep{CAMELS, CAMELS_public}. These  simulations can be classified into four different suites:
\begin{itemize}
    \item \textbf{IllustrisTNG}. Simulations run with the AREPO code \citep{Arepo} using the IllustrisTNG subgrid physics \citep{WeinbergerR_16a, PillepichA_16a}. 
    
    \item \textbf{SIMBA}. Simulations run with the GIZMO code \citep{Hopkins2015_Gizmo} using the SIMBA subgrid physics \citep{Dave2019_Simba}. 
    
    \item \textbf{Astrid}. Simulations run with the MP-Gadget code \citep{MPGadget} using the Astrid subgrid physics \citep{Astrid1, Astrid2}.

    \item \textbf{Magneticum}. Simulations run with the OpenGadget code using a similar but improved subgrid physics model following \citet{Fabjan2011MNRAS}, \citet{Hirschmann2014MNRAS}, \citet{Teklu} and \citet{Steinborn2016MNRAS}. 
\end{itemize}

Every suite contains 1,000 simulations (except Magneticum which contains 50 simulations), each of them with a different value of $\Omega_{\rm m}$, $\sigma_8$, $A_{\rm SN1}$, $A_{\rm SN2}$, $A_{\rm AGN1}$, and $A_{\rm AGN2}$ which are varied in a latin-hypercube with boundaries\footnote{In the case of the Astrid simulations, the parameter $A_{\rm AGN2}$ varies from 0.25 to 4.}:
\begin{eqnarray}
0.1\leq&\Omega_{\rm m}&\leq0.5\\
0.6\leq&\sigma_8&\leq1.0\\
0.25\leq &A_{\rm SN1}, A_{\rm AGN1}&\leq 4.0\\
0.5\leq &A_{\rm SN2}, A_{\rm AGN2}&\leq 2.0~.
\end{eqnarray}
The $A_{\rm SN}$ and $A_{\rm AGN}$ parameters control the efficiency of supernova and AGN feedback, and their specific definition depends on the considered subgrid model. We refer the reader to \citet{CAMELS} and \citet{CAMELS_Astrid} for further details on this. All simulations follow the evolution of $256^3$ dark matter particles plus $256^3$ initial fluid elements in a periodic box of $(25~h^{-1}{\rm Mpc})^3$ from $z=127$ down to $z=0$. We note that in all these simulations, the value of $\Omega_{\rm b}$ is fixed at 0.049.

Halos and subhalos are identified using \textsc{SUBFIND} \citep{Subfind, Dolag_2009} from all simulation snapshots. In this work, we consider galaxies\footnote{We define a galaxy as a subhalo with at least one star particle.}, with stellar masses $M_*\geq 5\times10^8~h^{-1}M_\odot$. We note that \citet{Cosmo1gal} considered galaxies with smaller stellar masses (e.g. galaxies with stellar masses $M_*\geq 2\times10^8~h^{-1}M_\odot$). We have checked that our conclusions do not change if we consider galaxies with smaller stellar masses. For each galaxy, we consider 14 properties, computed by \textsc{SUBFIND}:
\begin{enumerate}
\item \bm{$M_{\rm g}$}: The gas mass of the subhalo hosting the galaxy, including the contribution from the circumgalactic medium.
\item \bm{$M_{\rm BH}$}: The total mass of black-holes in the galaxy.
\item \bm{$M_*$}: The stellar mass of the galaxy.
\item \bm{$M_{\rm t}$}: The total mass of the subhalo hosting the galaxy.
\item \bm{$V_{\rm max}$}: The maximum circular velocity of the subhalo hosting the galaxy:  $V_{\rm max}=\max(\sqrt{GM(<R)/R}$).
\item \bm{$\sigma_v$}: The mass-weighted velocity dispersion of all particles contained in the galaxy's subhalo.
\item \bm{$Z_{\rm g}$}: The mass-weighted gas metallicity of the galaxy.
\item \bm{$Z_*$}: The mass-weighted stellar metallicity of the galaxy.
\item \bm{${\rm SFR}$}: The galaxy star formation rate.
\item \bm{$J$}: The modulus of the galaxy's subhalo spin vector.
\item \bm{$V$}: The modulus of the galaxy's subhalo peculiar velocity.
\item \bm{$R_*$}: The radius containing half of the galaxy's stellar mass.
\item \bm{$R_{\rm t}$}: The radius containing half of the total mass of the galaxy's subhalo.
\item \bm{$R_{\rm max}$}: The radius at which $\sqrt{GM(<R)/R}=V_{\rm max}$.
\end{enumerate}

We train three different models: 1) using IllustrisTNG galaxies, 2) using SIMBA galaxies, and 3) using Astrid galaxies. We then test the models on galaxies from all four suites. We note that we do not train a model on Magneticum galaxies as this suite only contains 50 simulations; not enough to train the models. 

We follow \cite{Cosmo1gal} and first split the simulations into training (900 simulations), validation (50 simulations), and testing (50 simulations). We then take individual galaxies from these simulations and pass them to the neural networks. In this way, we make sure that galaxies from one simulation are only been used for training, validation, or testing.

\subsection{Neural networks}

We train neural networks to infer the value of the cosmological and astrophysical parameters from the above 14 properties of individual galaxies. Our models consist of a series of fully connected layers with dropout and LeakyReLU activation functions. The number of layers, the number of neurons per layer, the value of the dropout, the weight decay, and the learning rate are hyperparameters that we optimize using Optuna \citep{Optuna}. 

Our models take as input a vector of 14 dimensions (representing the individual galaxy properties) and return $2N$ numbers, where $N$ is the number of parameters considered; $N=6$ when inferring all parameters and $N=1$ when only inferring $\Omega_{\rm m}$. For each parameter, the models predict the mean ($\mu_i$) and standard deviation ($\sigma_i$) of the marginal posterior for each parameter. To achieve this, we minimize the loss function of \citet{Jeffrey_2020} with the modifications described in \citet{CMD}. We emphasize that by construction, the output of our models represents the posterior mean and standard deviation without making any assumption about the shape of the posterior.

We perform more than 100 optuna trials\footnote{A trial represents a particular combination of the hyperparameter values.} minimizing the value of the validation loss. Unless stated explicitly, we only train our models on galaxies from a single simulation suite. However, we also investigate the behavior of our models when trained on galaxies from two different simulation suites (e.g. IllustrisTNG and Astrid).

\begin{figure*}[t]
\centering
\includegraphics[width=1.0\linewidth]{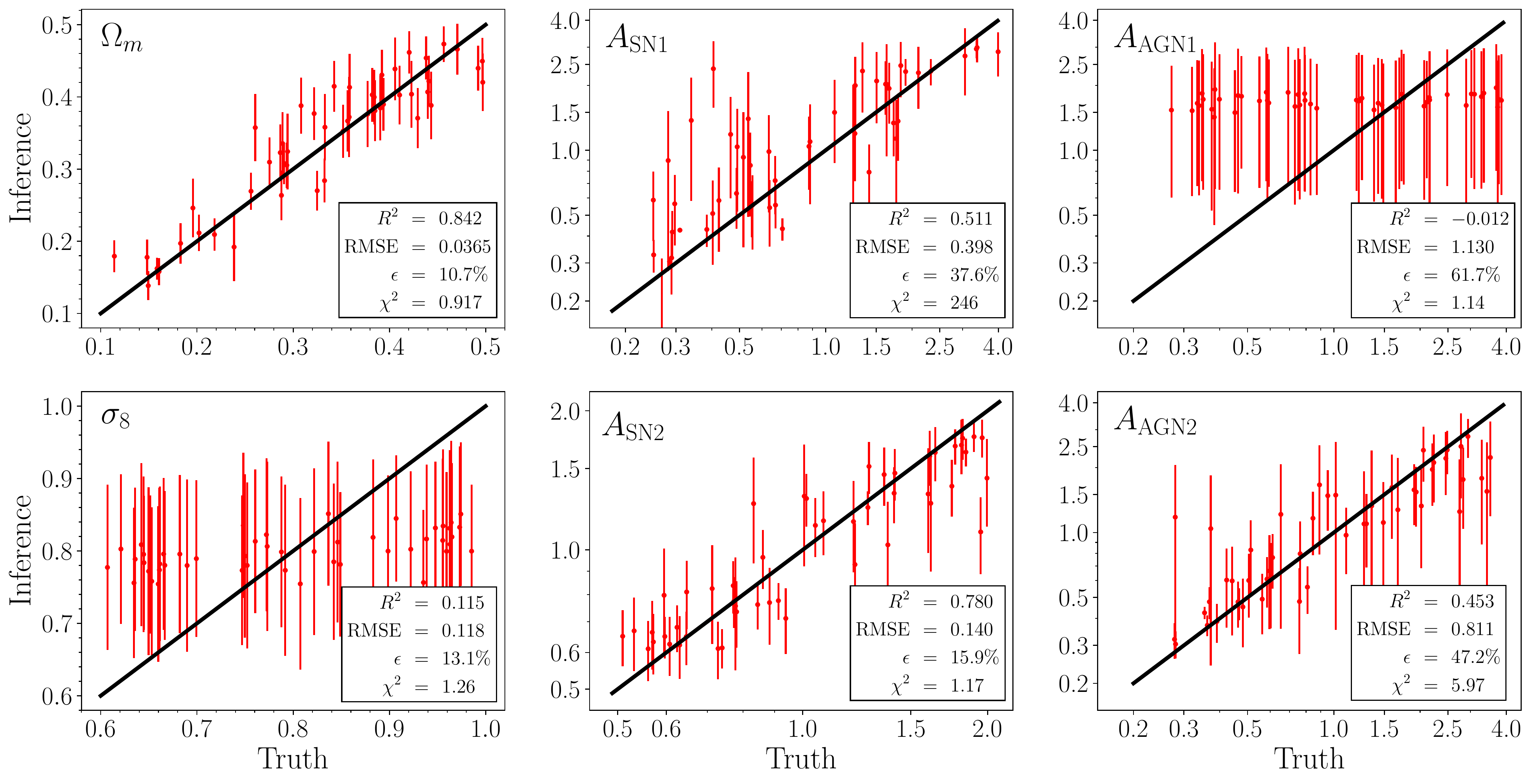}
\caption{We have trained a neural network to perform likelihood-free inference on the value of the cosmological ($\Omega_{\rm m}$ and $\sigma_8$) and astrophysical ($A_{\rm SN1}$, $A_{\rm SN2}$, $A_{\rm AGN1}$, and $A_{\rm AGN2}$) parameters using as input 14 properties of individual galaxies from the Astrid simulations at $z=0$. Once the network is trained, we test it using individual galaxies from the test set. The different panels show the posterior mean and standard deviation predicted by the network versus the true value. Every point with its error bar represents a single galaxy chosen randomly from each simulation of the test set. We find that our model is able to infer the value of $\Omega_{\rm m}$ from the properties of individual galaxies with a $\sim10\%$ precision.}
\label{fig:cosmo1gal_Astrid}
\end{figure*}

\subsection{Accuracy and precision metrics}

From the properties of a single galaxy, our models predict two numbers for the considered parameter $i$: the marginal posterior mean ($\mu_i$) and standard deviation ($\sigma_i$). We denote by $\theta_i$ the true value of the parameter $i$. In order to quantify the accuracy and precision of a given model, we made use of four different statistics:
\begin{enumerate}
\item \textbf{Root mean squared error} (RMSE), defined as 
\begin{equation}
{\rm RMSE}_i=\sqrt{\left\langle (\theta_i - \mu_i)^2 \right\rangle}~.
\end{equation}
The smaller the RMSE value the more accurate the model is.
\item \textbf{Mean relative error} ($\epsilon$), defined as
\begin{equation}
\epsilon_i = \left\langle \frac{\sigma_i}{\mu_i} \right\rangle~.
\end{equation}
The smaller the mean relative error, the more precise the model is.
\item \textbf{Coefficient of determination} ($R^2$), defined as 
\begin{equation}
    R_{i}^2 = 1 - \frac{\sum_i (\theta_i - \mu_i)^2}{\sum_i (\theta_i - \overline{\theta}_i)^2},
\end{equation}
The closer the value to 1 the more accurate the model is.
\item \textbf{Reduced chi-squared} ($\chi^2$), defined as 
\begin{equation}
    \chi_{i}^2=\frac{1}{N}\sum_{i=1}^N \left(\frac{\theta_i-\mu_i}{\sigma_i}\right)^2~.
\end{equation}
The value of the reduced $\chi^2$ is used to quantify the reliability of the errors (posterior standard deviation for us). Values close to one indicate the errors are properly quantified, while values larger/smaller than one show that the errors are underestimated/overestimated.
\end{enumerate}

\section{Results}
\label{sec:results}

In this section, we first present the results obtained by training models on Astrid galaxies. We then study the robustness of this and the models trained on IllustrisTNG and SIMBA galaxies.

\subsection{Astrid galaxies}

\citet{Cosmo1gal} showed that neural networks were able to infer the value of $\Omega_{\rm m}$ from individual galaxies from either IllustrisTNG or SIMBA simulations. Here we investigate whether this claim holds for individual galaxies generated by a completely different code (MP-Gadget) that uses an independent and different subgrid model (Astrid).

We train our model on properties from individual galaxies of the Astrid simulations at $z=0$ to infer the value of all parameters ($\Omega_{\rm m}$, $\sigma_8$, $A_{\rm SN1}$, $A_{\rm SN2}$, $A_{\rm AGN1}$, and $A_{\rm AGN2}$). We then test the model on individual galaxies from Astrid but coming from simulations whose galaxies the model has never seen. We show the results of this test in Fig. \ref{fig:cosmo1gal_Astrid}. 

As can be seen, the model is able to infer the value of $\Omega_{\rm m}$ from the properties of individual galaxies with high accuracy and precision. Similarly to \citet{Cosmo1gal} we find that the model is unable to infer the value of $\sigma_8$ and $A_{\rm AGN1}$. On the other hand, our model seems to be able to infer the value of $A_{\rm SN1}$, $A_{\rm SN2}$, and $A_{\rm AGN2}$, although with large error bars. The accuracy and precision metrics for each parameter are reported in the bottom-right part of each panel.

In the case of $\Omega_{\rm m}$, the mean relative error is $\sim11\%$ and $R^2=0.842$, indicating good precision and accuracy. The value of $\chi^2$ is close to 1, showing that the error bars are accurately estimated. We note that some parameters have a very large value of $\chi^2$ (e.g. $A_{\rm SN1}$). This is due to a few outliers that contribute largely. Overall, the accuracy and precision metrics indicate that the model is well-trained.

From now on we will focus our attention on inferring the value of $\Omega_{\rm m}$ and leave all other parameters aside. In order to improve the precision and accuracy of the model, we retrain to predict only the posterior mean and standard deviation of $\Omega_{\rm m}$. We do this because this is usually an easier task than inferring several parameters at the same time; a more complicated task prone to degeneracies and local minima.

\begin{figure*}
\centering
\includegraphics[width=1.0\linewidth]{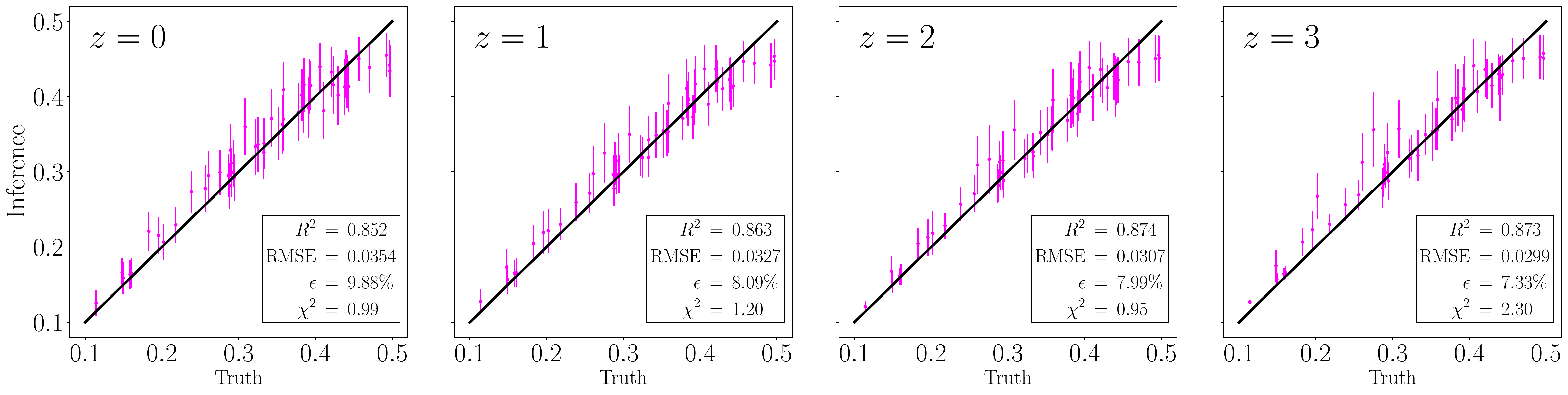}
    \caption{Redshift dependence: We have trained neural networks to infer the value of $\Omega_{\rm m}$ using properties of individual galaxies at different redshifts and for galaxies of the Astrid simulations. For each galaxy of each simulation of the test set we compute the posterior mean and standard deviation for $\Omega_{\rm m}$. Next, we compute the mean of those two numbers (Eq. \ref{Eq:mean_values}) and plot them in the figure for the 50 different simulations in the test set. We show results at redshifts 1, 2, and 3. As can be seen, our networks can infer the value of $\Omega_{\rm m}$ from individual galaxies at redshifts higher than $z=0$ with an accuracy similar to the one achieved by the models at $z=0$.}
    \label{fig:redshift_evolution}
\end{figure*}

It is interesting to visualize the average results for all galaxies in a given simulation, rather than individual galaxies. In this way, we are less sensitive to outliers and we can detect biases more easily. To carry out this task, we compute 
\begin{equation}
\bar{\mu}_i = \frac{1}{N_s}\sum_{j\in s} \mu_{i,j} \hspace{1cm} \bar{\sigma}_i = \frac{1}{N_s}\sum_{j\in s} \sigma_{i,j}~,
\label{Eq:mean_values}
\end{equation}
where $i$ denotes the considered parameter (e.g. $\Omega_{\rm m}$) and $j$ runs over all $N_s$ galaxies of a given simulation $s$. We show the average results for all 50 simulations in the test set in the left panel of Fig. \ref{fig:redshift_evolution}. 

From the metrics, we can see that this model is indeed slightly more accurate and precise than the one used to infer all 6 properties. Overall, we see that the model is able to infer the value of $\Omega_{\rm m}$ with a small bias in most of the cases. From these results, we can already conclude that Astrid galaxies also exhibit a tight relationship between $\Omega_{\rm m}$ and their individual properties. We emphasize that this relation, as determined by the networks, already accounts for changes in supernova and AGN parameters as modeled in CAMELS.

\subsubsection{Redshift dependence}

We now study whether the tight relation between $\Omega_{\rm m}$ and galaxy properties holds at redshifts other than $z=0$. For this, we train our models on Astrid galaxies at redshifts 1, 2, and 3 and compute the mean values of all galaxies in a given simulation according to Eq. \ref{Eq:mean_values}. We then show the results in Fig. \ref{fig:redshift_evolution}. 

As can be seen, our models perform well at all considered redshifts. Our results indicate a tighter relation between $\Omega_{\rm m}$ and galaxy properties at higher redshifts. This could be due to astrophysics effects being less severe on galaxy properties. We note that these results are in agreement with those of \citet{Cosmo1gal}, who performed a similar analysis for IllustrisTNG and SIMBA galaxies. We note that a model trained on galaxies at a given redshift will not work if tested on galaxies at a different redshift, also in agreement with the results of \cite{Cosmo1gal}.

\subsubsection{Relevant features}

We now investigate what are the most relevant galaxy properties used by the models to infer the value of $\Omega_{\rm m}$. For this, we follow the procedure utilized in \citet{Cosmo1gal} that we briefly describe here. First, a gradient-boosted tree regressor\footnote{We use this method instead of neural networks as this task would be too computationally expensive to carry out with neural networks.} is used to predict the value of $\Omega_{\rm m}$ from the 14 properties of individual galaxies. Next, one of the galaxy properties is removed from the input, the regressor is retrained, and its accuracy is saved. This procedure is repeated for all 14 properties. The set with the 13 properties that achieves the highest accuracy is kept for the next phase, and the property outside that set is discarded. The above procedure is then repeated by removing one property at a time until the set only contains one property.

\begin{figure*}[t]
\centering
\includegraphics[width=0.7\linewidth]{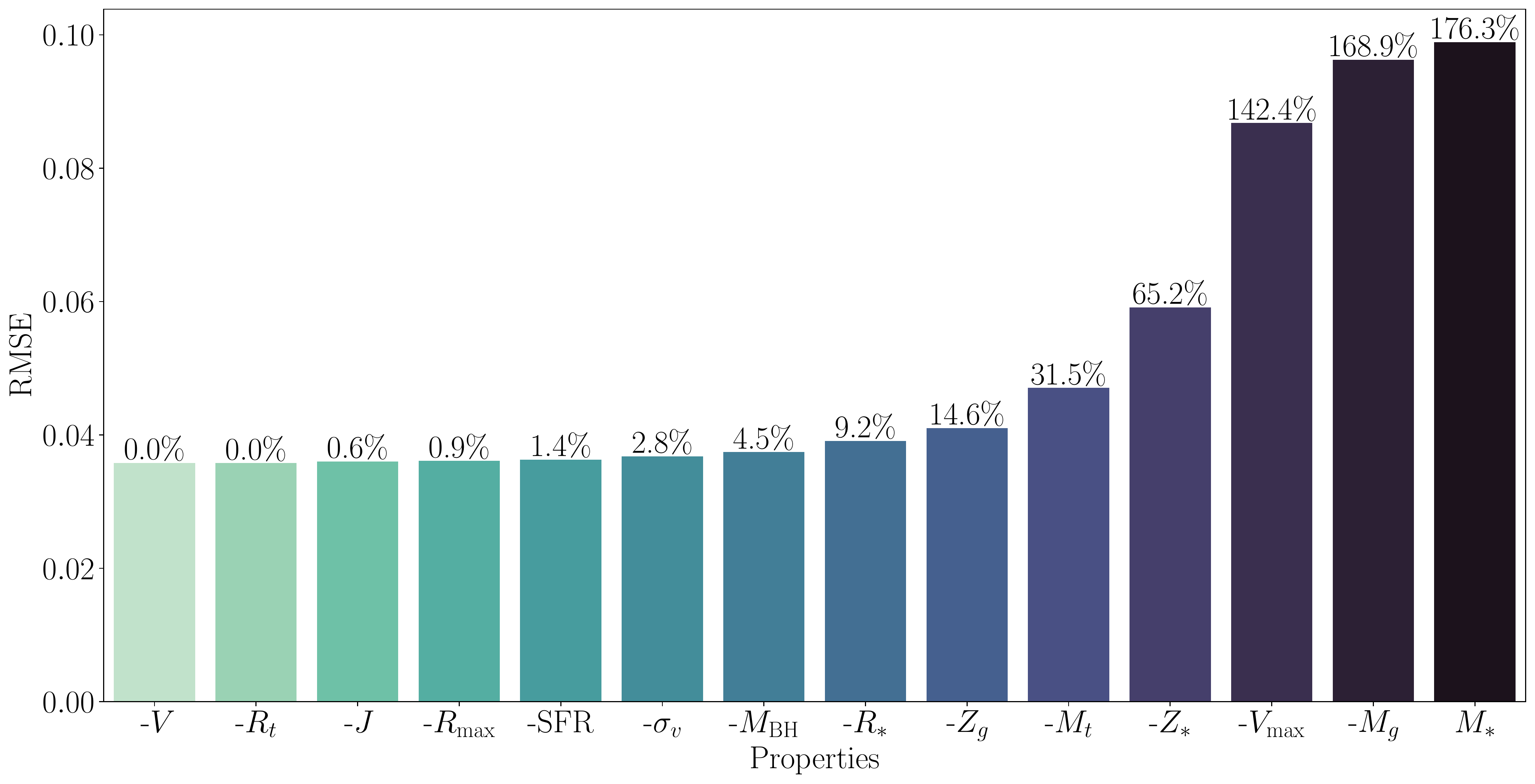}
\caption{We rank order the galaxy properties for Astrid such that the variables contributing the most to the model accuracy are on the right while the features contributing the least are on the left (see text for details on the procedure used). The vertical bars indicate the accuracy (in terms of RMSE) achieved by the considered variables, cumulatively from left to right, and the black numbers inside them show the loss in accuracy with respect to a model trained using all variables. For instance, a model that only uses $M_{\rm *}$ achieves an RMSE of $\sim0.1$ and performs 176.3\% worse than the model trained on all 14 properties (with an RMSE of $\sim0.04$).We emphasize that this ordering was derived when training gradient boosting trees models to perform regression to the value of $\Omega_{\rm m}$.}
\label{fig:ranking}
\end{figure*}

\begin{figure*}
\centering
\includegraphics[width=1.0\linewidth]{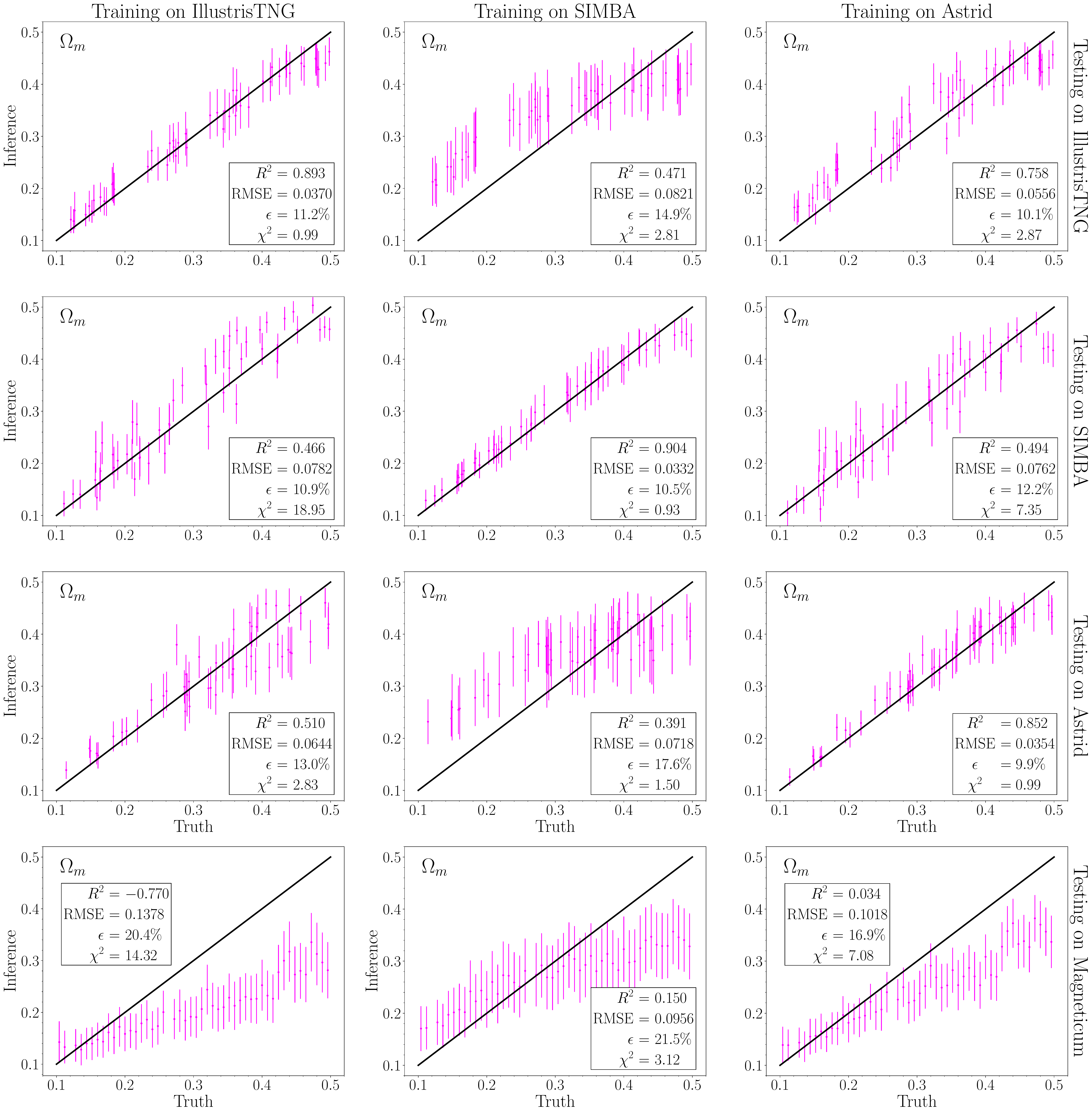}
\caption{Robustness test. We test models trained on individual galaxies from IllustrisTNG (left column), SIMBA (middle column), and Astrid (right column) on galaxies from the IllustrisTNG (first row), SIMBA (second row), Astrid (third row), and Magneticum (fourth row) suites. Each point represents the average result of all galaxies in that simulation (see Eq. \ref{Eq:mean_values}). We find that none of the models are robust. On the other hand, the model trained on Astrid galaxies performs relatively well when tested on IllustrisTNG and SIMBA galaxies.}
\label{fig:robustness_no_chi2_cut}
\end{figure*}

The above procedure\footnote{We note that this procedure is not meant to be optimal, and other sets of variables may yield similar results.} allow us to identify sets of variables that carry different fractions of the information. In Fig. \ref{fig:ranking} we show the loss in accuracy of the model as we discard galaxy properties. For instance, keeping all galaxy properties but peculiar velocities and $R_t$ have a negligible effect on the accuracy of the model. On the other hand, the set $\{ M_*, M_g, V_{\rm max}, Z_*, M_t\}$ yields an accuracy $\sim30\%$ worse than the model trained on all galaxy properties. Adding one more variable, $\{ M_*, M_g, V_{\rm max}, Z_*, M_t, Z_g\}$, further improves the accuracy: only $\sim15\%$ worst than using all galaxy properties.

It is very interesting to see that three of the most relevant galaxy properties are 1) the stellar mass ($M_*$), 2) the maximum circular velocity ($V_{\rm max}$), and 3) the stellar metallicity ($Z_*$). Those variables are also among the most relevant for galaxies in the IllustrisTNG and SIMBA models \citep{Cosmo1gal}.

\subsection{Robustness}
\label{subsec:robust}

It is important to investigate how well the different models generalize; i.e. whether the networks are able to infer the value of $\Omega_{\rm m}$ from galaxies from simulations run with different codes to the ones used for training. \citet{Cosmo1gal} showed that the models trained on IllustrisTNG galaxies were not able to infer the correct value of $\Omega_{\rm m}$ when tested on SIMBA galaxies (and the other way around).

We have trained three different models using $z=0$ galaxies from 1) IllustrisTNG, 2) SIMBA, and 3) Astrid simulations. We then test the models on galaxies from all four codes: IllustrisTNG, SIMBA, Astrid, and Magneticum. To simplify the analysis, we compute the mean results from all galaxies using Eq. \ref{Eq:mean_values}. We show the results in Fig. 
\ref{fig:robustness_no_chi2_cut}. We emphasize that the performance metrics shown in the different panels represent the results of taking the average over all galaxies in the test set; for instance, $\chi^2=(\sum_i^N \chi^2_i)/N$.
 
First, we are able to reproduce the results of \citet{Cosmo1gal} and the models trained on IllustrisTNG/SIMBA galaxies do not perform well when tested on SIMBA/IllustrisTNG galaxies\footnote{Note that in this case we are using a slightly different cut in stellar mass when selecting the galaxies, so results are similar but not identical.}. On top of this, we find that those models do not perform well when tested on galaxies from Astrid and Magneticum simulations. Similarly, we find that the model trained on Astrid galaxies does not perform well when tested on IllustrisTNG, SIMBA, and Magneticum galaxies. It is interesting to see that the models trained on IllustrisTNG (Astrid) galaxies do not perform that badly when tested on Astrid (IllustrisTNG) galaxies, perhaps signaling similarities between these two simulations.

\begin{figure*}
\centering
\includegraphics[width=0.95\linewidth]{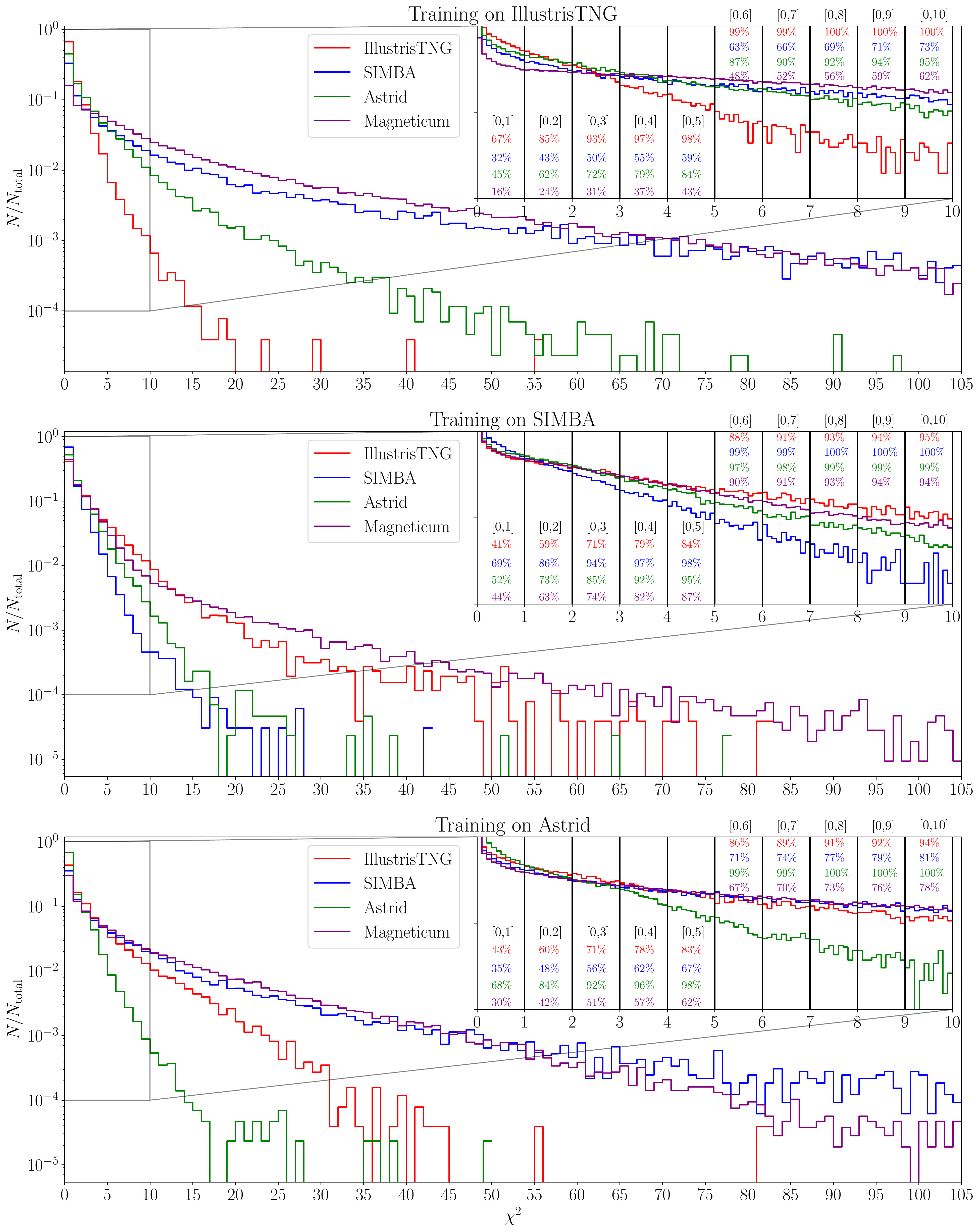}
\caption{We have trained three different models using IllustrisTNG (top), SIMBA (middle), and Astrid (bottom) galaxies at $z=0$. We test each model on galaxies from the IllustrisTNG, SIMBA, Astrid, and Magneticum simulations. For each galaxy $i$ we compute $\chi^2_i=(\theta_i-\mu_i)^2/\sigma_i^2$. The different lines show the distribution of the $\chi^2$ in the different setups. As can be seen, the $\chi^2$ distribution changes if the simulation is tested on galaxies from a different code than the one used for training. On the other hand, most of the galaxies have low $\chi^2$ values. The numbers in the subpanels indicate the fraction of galaxies with values of $\chi^2$ smaller than the indicated threshold.}
\label{fig:chi2_distribution}
\end{figure*}

\begin{figure*}
\centering
\includegraphics[width=1.0\linewidth]{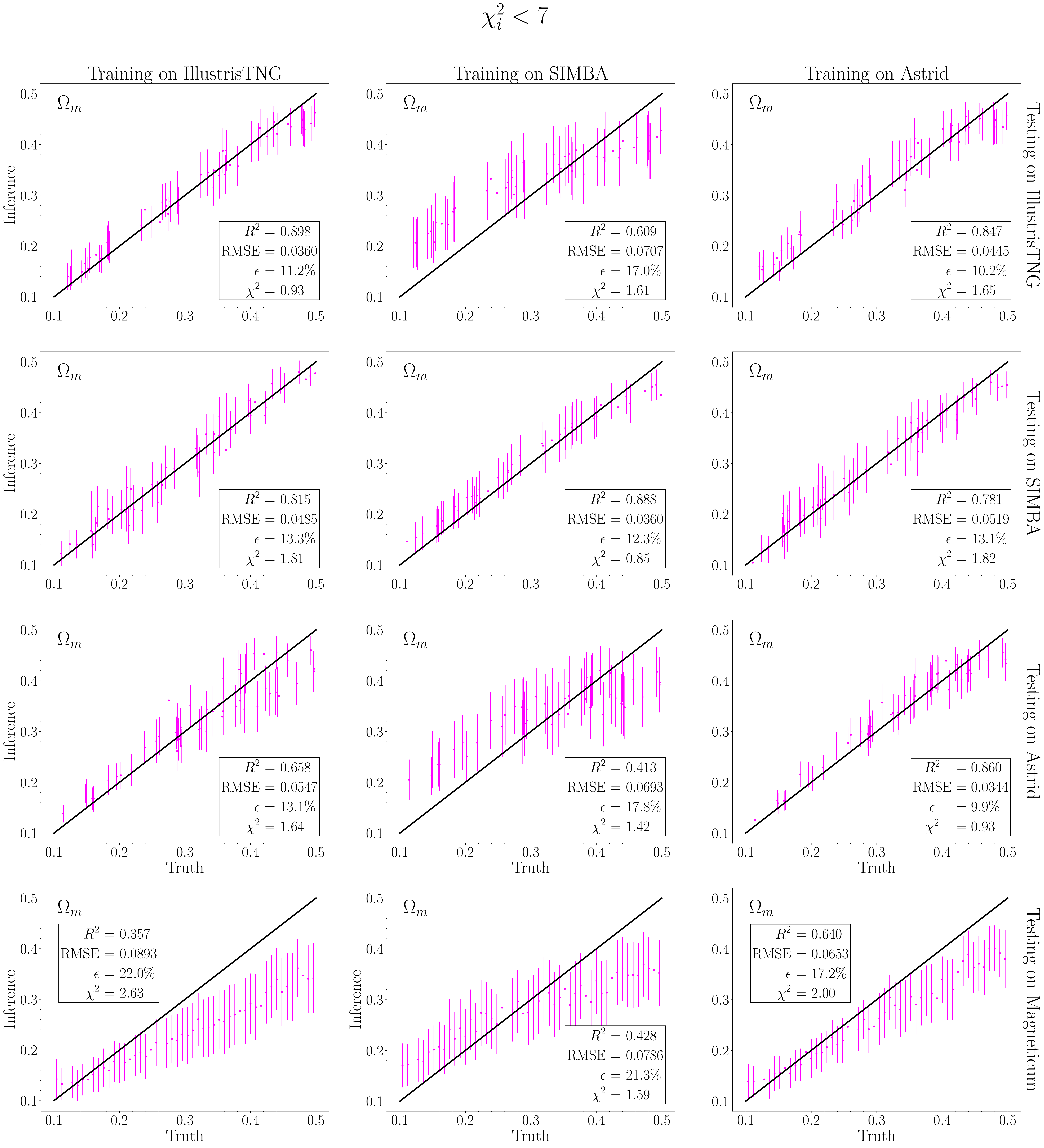}
\caption{Same as Fig. \ref{fig:robustness_no_chi2_cut} but removing all galaxies whose individual $\chi^2$ values are higher than 7. As can be seen, the models trained on Astrid and IllutrisTNG galaxies are very robust (with the exception of Magneticum galaxies with $\Omega_{\rm m}\gtrsim0.3$) while the model trained on SIMBA galaxies is not.}
\label{fig:robustness_chi2_cut}
\end{figure*}

We note that, from Fig. \ref{fig:robustness_no_chi2_cut}, we can only reach conclusions about the mean behavior of the models. Thus, there are different possibilities that can explain our results. First, it could be that the models fail because galaxies from different codes are very different; in this case, we would expect a generic failure of the model. In other words, the networks should infer wrong values of $\Omega_{\rm m}$ for all (or the majority) of galaxies. Second, it could be that the mean of the models is off due to the presence of some outliers where the models fail catastrophically. In order to shed light on this, we have computed, for each individual galaxy $i$ in the test set, the value of its reduced chi-squared
\begin{equation}
\chi^2_i=\frac{(\theta_i-\mu_i)^2}{\sigma^2_i}~,
\end{equation}
where $\theta_i$ is the value of $\Omega_{\rm m}$ of the galaxy while $\mu_i$ and $\sigma_i$ are the posterior mean and standard deviation predicted by the network. 

In Fig. \ref{fig:chi2_distribution} we show the distribution of the $\chi^2$ values for the individual galaxies of the test set of the different simulations. We find that most galaxies have low $\chi^2$ values in all cases. For instance, 83\%, 67\%, 98\%, and 62\% of the IllustrisTNG, SIMBA, Astrid, and Magneticum galaxies have $\chi^2$ values below 5 when tested on the model trained on Astrid galaxies. However, the $\chi^2$ distribution for galaxies tested on models different from the ones used for training display long tails with large $\chi^2$ values; $27\%$, $5\%$, and $38\%$ of the SIMBA, Astrid, and Magneticum galaxies have $\chi^2$ values larger than 10 when tested on the model trained on IllustrisTNG galaxies.

This indicates that the failure of the models is due to the presence of outliers. To verify this, we have removed all galaxies with $\chi_i^2>7$ from the test sets. We emphasize that our models are trained using all galaxies in the considered suite (e.g. all galaxies in the IllustrisTNG training set). We then compute the mean values of all remaining galaxies using Eq. \ref{Eq:mean_values} and show the results in Fig. \ref{fig:robustness_chi2_cut}. 

As can be seen, the performance metrics of all models significantly improve. In particular, the models trained on IllustrisTNG or Astrid galaxies perform well on galaxies from all simulations, with the exception of Magneticum galaxies with large values of $\Omega_{\rm m}$. It is also interesting to note that although the model trained on SIMBA galaxies exhibits low $\chi^2$ values, it is not robust: for low values of $\Omega_{\rm m}$ it systematically overpredicts the true value. On the other hand, this model performs better on Magneticum galaxies than the other two models.

We emphasize that removing outliers will naturally lead to better predictions overall, so it is not surprising that the models become more robust when using this method. On the other hand, it is important to note that only a relatively small fraction of the galaxies behave as outliers. For instance, for the model trained on Astrid galaxies, we find that only $11\%$, $26\%$, $<1\%$, and $30\%$ of the IllustrisTNG, SIMBA, Astrid, and Magneticum galaxies have $\chi^2_i \geq 7$. 

Overall, the model trained on Astrid galaxies seems to be one with the best generalization properties once the outliers are removed. The fact that it fails for Magneticum galaxies with large values of $\Omega_{\rm m}$ may be due to intrinsic differences between Magneticum simulations and the other models \cite[see e.g.][]{deSanti_2023}. We have checked that more aggressive cuts in the $\chi^2_i$ lead to less unbiased predictions, as expected (see Appendix \ref{sec:chi2_cut}). Similar conclusions can be reached for SIMBA galaxies: by applying more aggressive cuts in the $\chi^2$, the models become more and more robust across simulations. On the other hand, we also know that Astrid galaxies exhibit the most diverse set of properties \citep{CAMELS_Astrid}. Thus, it is perhaps expected that they perform best when tested on galaxies from other simulations.

These results indicate that we could develop robust models simply by knowing their range of validity; in other words, by not using them in cases where they will not perform well. A natural question arises: how do we identify model outliers a-priori? In other words, if we are not able to compute the $\chi^2$ of a galaxy (e.g. with a real galaxy, we do not know the true value of $\Omega_{\rm m}$), how can we flag it as an outlier? While we do not provide a rigorous answer to this question in this paper, we have investigated the distribution of the galaxy properties for these outliers. We find evidence suggesting that those outliers correspond to galaxies whose properties are far away from the distribution of the galaxies used for training. We provide details on this test in Appendix \ref{sec:outliers}. These results indicate that the outliers of the model may simply correspond to outliers in the distribution. 

We conclude this section by noting that training models on galaxies from two different codes do not seem to help in improving the robustness of the model. This was one of the hypotheses suggested by \cite{Cosmo1gal} to improve the generalization capabilities of the models. We provide further details on this in  Appendix \ref{sec:two_suites}.

\section{Conclusions}
\label{sec:summary}

We now summarize the main takeaways from this work:

\begin{itemize}

    \item \citet{Cosmo1gal} claimed that it is possible to infer the value of $\Omega_{\rm m}$, at fixed $\Omega_{\rm b}$, from the properties of individual simulated galaxies. They showed that their models were able to perform that task using galaxies generated by IllustrisTNG and SIMBA simulations. In this work, we have shown that it is also possible to infer $\Omega_{\rm m}$, at fixed $\Omega_{\rm b}$, with a $\sim10\%$ precision from properties of individual galaxies generated by Astrid simulations that employ a different method to solve the hydrodynamic equations and utilize a completely different subgrid model than IllustrisTNG and SIMBA simulations.
    
    \item The properties of the Astrid galaxies seem to be more sensitive to the value of the astrophysical parameters than the IllustrisTNG and SIMBA galaxies. Because of this, our models are able to infer the value of 
    $A_{\rm SN2}$ and $A_{\rm AGN2}$ (although with large error bars); this was not possible with IllustrisTNG and SIMBA galaxies.
    
    \item The tight relation between $\Omega_{\rm m}$ and the properties of individual Astrid galaxies is present at all redshifts considered in this work: $z=0, 1, 2$, and $3$. Models trained at higher redshifts are able to infer $\Omega_{\rm m}$ slightly more accurately.
    
    \item The five more important properties used by the model to infer $\Omega_{\rm m}$ from Astrid galaxies are $\{ M_*,M_g,V_{\rm max}, Z_*, M_t\}$. By using only these properties, our models are able to infer $\Omega_{\rm m}$ with an accuracy that is only $30\%$ worst than when using all 14 galaxy properties. Interestingly, the stellar mass, the maximum circular velocity, and the stellar metallicity are top properties for models trained on IllustrisTNG, SIMBA, or Astrid simulations.
    
    \item The model trained on Astrid galaxies is not robust, and it fails when tested on IllustrisTNG, SIMBA, and Magneticum galaxies. The models trained on IllustrisTNG and SIMBA galaxies also perform badly when tested on Astrid and Magneticum galaxies.
    
    \item An important factor behind the lack of robustness in our models is the presence of outliers. As expected, removing those outliers significantly improves the robustness of the models. We note that the fraction of outliers is relatively small. For instance, for the model trained on Astrid galaxies, only $11\%$, $26\%$, $<1\%$, and $30\%$ of the IllustrisTNG, SIMBA, Astrid, and Magneticum galaxies have $\chi^2\geq7$.

    \item We note that all models exhibit a bias when tested on Magneticum galaxies with $\Omega_{\rm m}\gtrsim 0.3$; even after applying the $\chi^2_i>7$ cut. This bias can be due to the fact that those simulations exhibit systematic differences with respect to others \citep{deSanti_2023} models, and therefore may require a more aggressive cut. We have checked that more aggressive cuts improve the performance of the model.

    \item Our results indicate that model outliers (defined as galaxies with $\chi^2_i\geq7$) tend to correspond to galaxies with properties either outside or in the tails of the galaxy distribution (see Appendix \ref{sec:outliers}).

    \item Training on galaxies from two different simulations (e.g. IllustrisTNG and SIMBA) does not make the model robust and it still fails when testing it on a third simulation.
\end{itemize}

It is important to emphasize that we identify outliers as galaxies having large $\chi^2$ values ($\chi_i^2\geq7$). Removing these outliers will naturally decrease the mean $\chi^2$ value of the whole population, so it is not surprising that our models become more robust after performing this task. The important thing to note is that those outliers only represent a relatively small fraction of the galaxies. 

Identifying outliers as galaxies with large $\chi_i^2$ values can only be done if the true value of $\Omega_{\rm m}$ is known. Thus, this method cannot be used with real galaxies. On the other hand, we have some hints that outliers tend to correspond to galaxies whose properties reside on the outskirts of the distribution used for training (see Appendix \ref{sec:outliers}). Therefore, it may be possible to identify outliers by finding galaxies whose properties are far away from the manifold that contains galaxy properties. Being able to identify and discard outliers will improve the robustness of the model as we have shown in this paper. This opens the door to being able to apply our method to real data if the galaxies we consider are not an outlier with respect to those we train the models on. We will investigate in detail this avenue in future work. 

Finally, we emphasize that a model that is able to infer the value of $\Omega_{\rm m}$ from galaxies of simulations run with three different codes, is not guaranteed to perform well on simulations from a new simulation. This is clearly illustrated with the Magneticum galaxies; even after removing the outliers, the model trained on Astrid galaxies does not perform well on Magneticum galaxies. In this case, we may need to be even more aggressive in the way we define outliers to improve the robustness of the models. 

Overall, in this work we have shown that galaxies from three different types of simulations (run with different codes and employing different subgrid physics models) exhibit a tight relationship between $\Omega_{\rm m}$ and their individual internal properties. This relation is not affected by astrophysics (at least in the way we model them in CAMELS) since our simulations vary astrophysical parameters controlling the efficiency of supernova and AGN feedback. While our models are not robust yet, in this work we have shown that identifying and removing outliers seems a promising way to address this issue. This method may also make the models robust to effects from supersample covariance and changes in astrophysics and cosmological parameters not covered in CAMELS.

\section*{ACKNOWLEDGEMENTS}
We have made use of the XGB\footnote{\url{https://xgboost.readthedocs.io}}, PyTorch, and Optuna packages. We thank Marc Huertas-Companys, Arnab Lahiry, Natali de Santi, and Helen Shao for useful conversations. We thank the RECA\footnote{\url{https://www.astroreca.org/en}} (Red Estudiantes Colombianos en Astronom\'ia) Summer Internship Program for providing an avenue to initiate the discussions that led to this work. NE thanks the Simons Foundation for support while carrying out this work. The work of FVN is supported by the Simons Foundation. The CAMELS project is supported by the Simons Foundation and the NSF grant AST 2108078. DAA acknowledges support by NSF grants AST-2009687 and AST-2108944, CXO grant TM2-23006X, Simons Foundation Award CCA-1018464, and Cottrell Scholar Award CS-CSA-2023-028 by the Research Corporation for Science Advancement. KD acknowledges support by the COMPLEX project from the European Research Council (ERC) under the European Union’s Horizon 2020 research and innovation program grant agreement ERC-2019-AdG 882679 as well as support by the Deutsche Forschungsgemeinschaft (DFG, German Research Foundation) under Germany’s Excellence Strategy - EXC-2094 - 390783311. Details on the CAMELS simulations can be found in \url{https://www.camel-simulations.org}.

\appendix 

\section{$\chi^2$ cut}
\label{sec:chi2_cut}

\begin{figure*}
\centering
\includegraphics[width=1.0\linewidth]{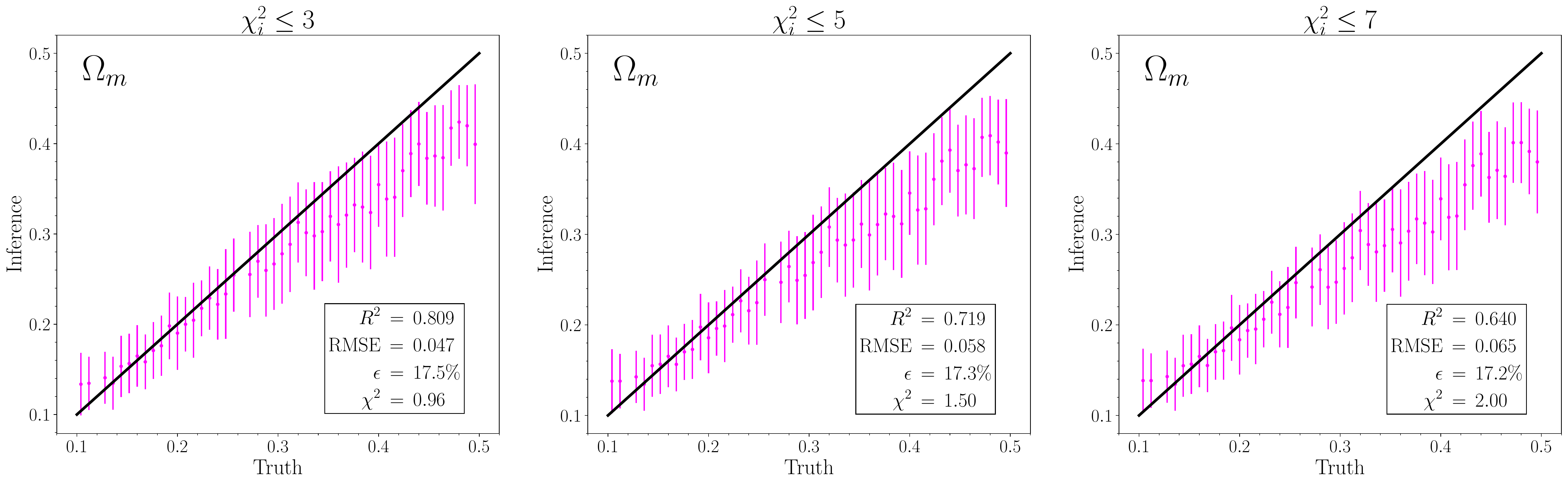}
\caption{We have tested the model trained on Astrid galaxies on Magneticum galaxies after removing galaxies with $\chi^2_i\geq3$ (left), $\chi^2_i\geq5$ (middle), and $\chi^2_i\geq7$ (right). As expected, the more aggressive we are in removing outliers the better the model works.}
\label{fig:chi2_cut}
\end{figure*}

In Fig. \ref{fig:robustness_chi2_cut} we saw that all models failed when tested on Magneticum galaxies with $\Omega_{\rm m}\gtrsim0.3$ even after removing the galaxies with $\chi^2_i\geq7$. We discussed that more aggressive cuts could improve the robustness of the model. In order to verify that, we have tested the model trained on Astrid galaxies on Magneticum galaxies after removing galaxies with $\chi^2_i$ greater than 3, 5, and 7. We show the results in Fig. \ref{fig:chi2_cut}. As expected, the model becomes more robust the more aggressive our cuts are. We reach similar conclusions in other scenarios, e.g. training on SIMBA galaxies and testing on IllustrisTNG and Astrid galaxies.

\begin{figure*}[h]
\centering
\includegraphics[width=1.0\linewidth]{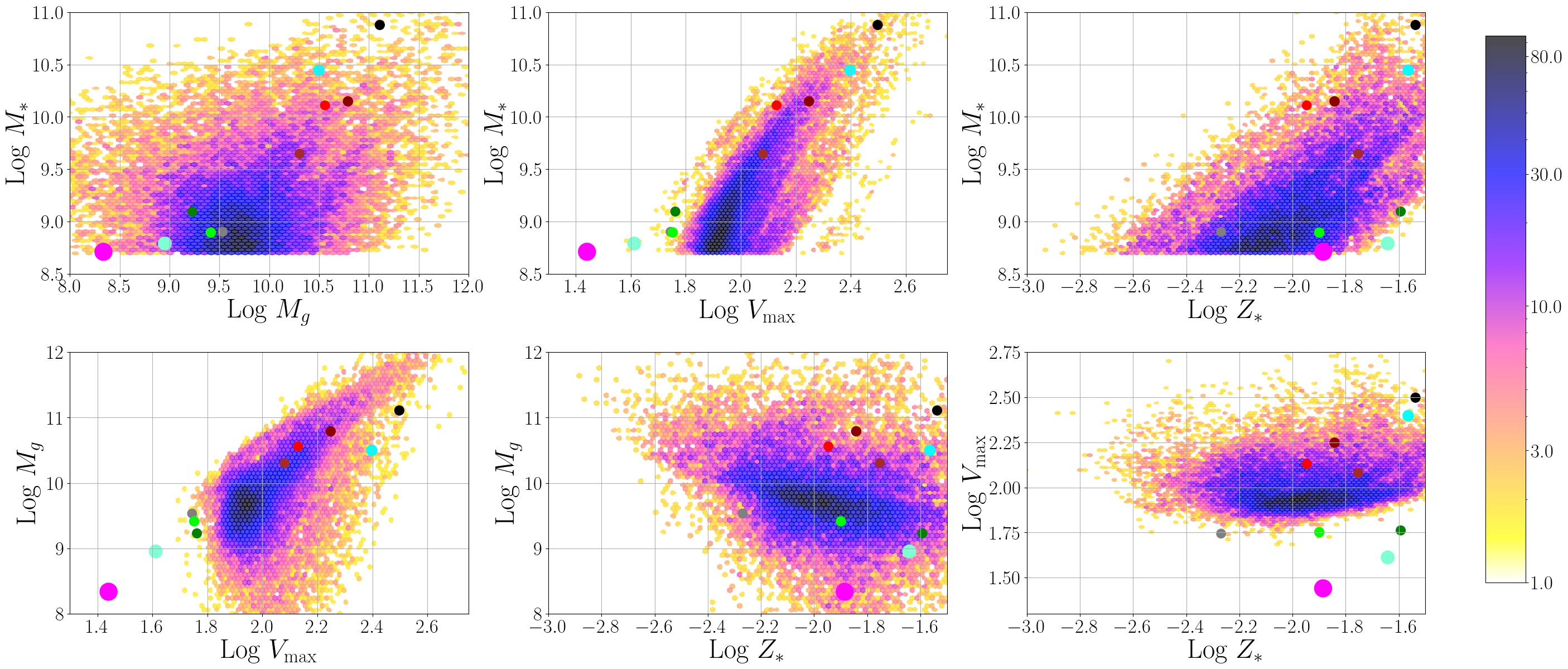}
\caption{For each galaxy in the training set of the Astrid suite we have considered four properties: 1) stellar mass, $M_*$, 2) gas mass, $M_g$, 3) maximum circular velocity, $V_{\rm max}$, and 4) stellar metallicity, $M_*$. The different panels show the 2D distribution of these properties with  hexagons. The color of the hexagons indicates the number of galaxies in that region of parameter space (see color bar). We then test our model, trained on Astrid galaxies, on IllustrisTNG galaxies. For each of those galaxies we compute their $\chi^2_i$ value and select the 10 galaxies with the highest values. The $\chi^2_i$ of these galaxies ranges from 40 to 144.
We then project these galaxies into the different 2D properties. The colors of the points are used to match the galaxies across panels. As can be seen, these galaxies tend to reside in regions in parameter space with very low density in one or several dimensions. This indicates that these outliers tend to correspond to galaxies whose properties are on the tails (or outside) of the Astrid distribution.}
\label{fig:properties_Illustris}
\end{figure*}

\section{Outliers}
\label{sec:outliers}

\begin{figure*}[h]
\centering
\includegraphics[width=1.0\linewidth]{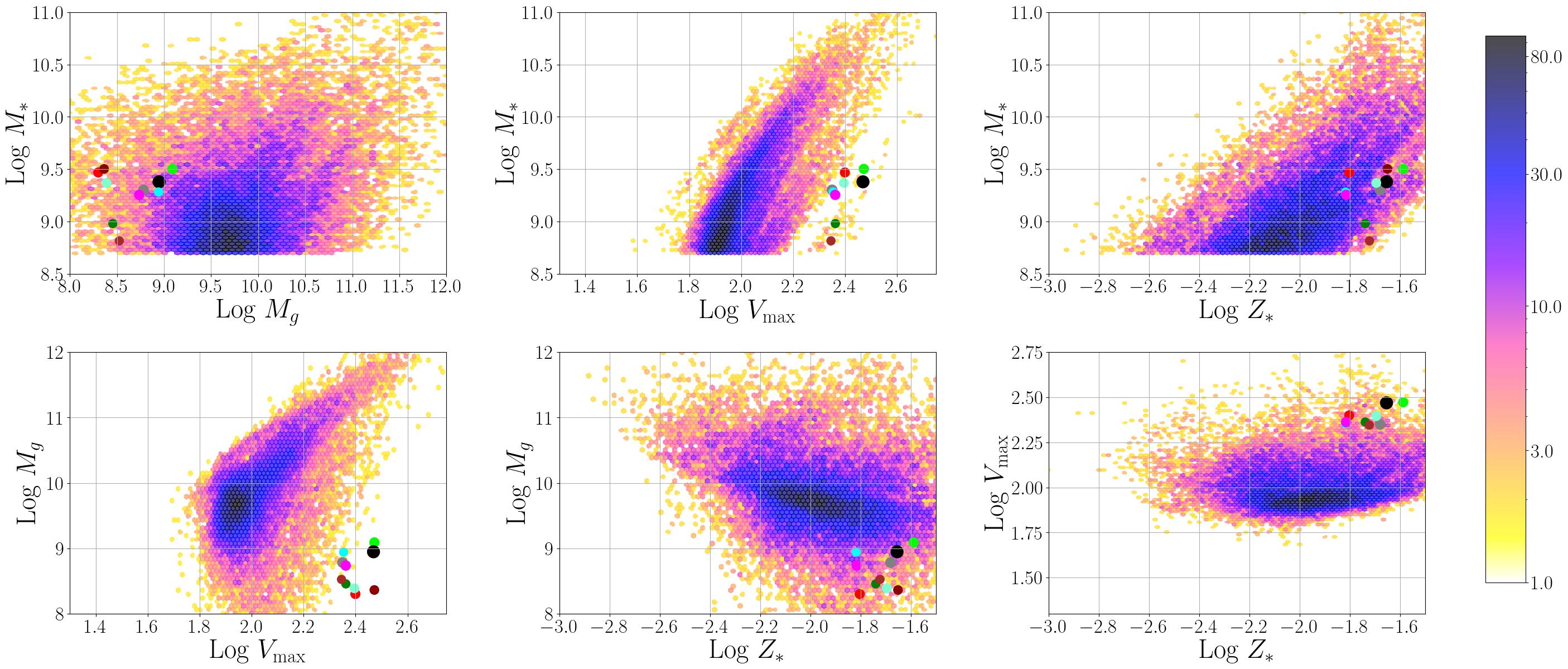}
\caption{Same as Fig. \ref{fig:properties_Illustris} but testing the model on SIMBA galaxies. The $\chi^2_i$ of these galaxies range from 252 to 531.}
\label{fig:properties_SIMBA}
\end{figure*}

\begin{figure*}[h]
\centering
\includegraphics[width=1.0\linewidth]{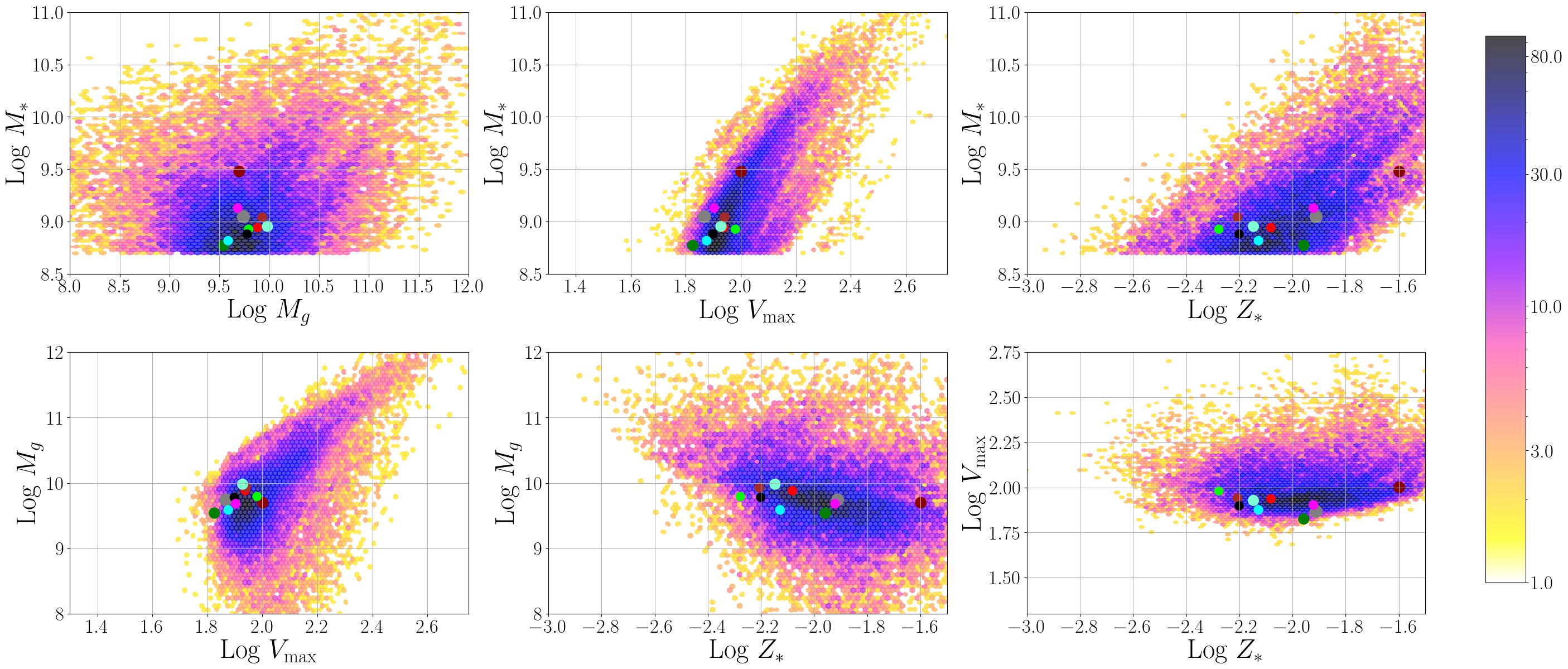}
\caption{Same as Fig. \ref{fig:properties_Illustris} but testing the model on Astrid galaxies. The $\chi^2_i$ of these galaxies ranges from 24 to 49.}
\label{fig:properties_astrid}
\end{figure*}

\begin{figure*}[h]
\centering
\includegraphics[width=1.0\linewidth]{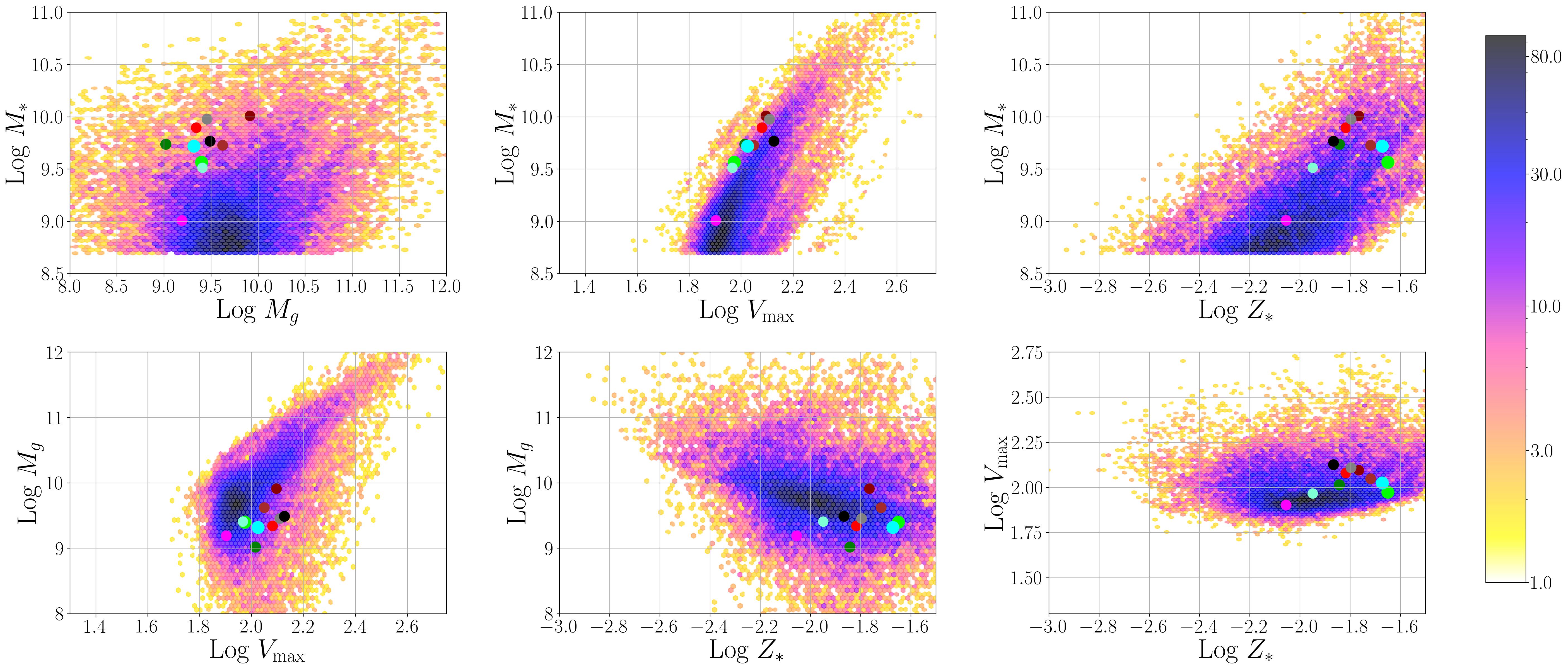}
\caption{Same as Fig. \ref{fig:properties_Illustris} but testing the model on Magneticum galaxies. The $\chi^2_i$ of these galaxies ranges from 111 to 180.}
\label{fig:properties_magneticum}
\end{figure*}

Here we study whether the model outliers, defined as galaxies with $\chi^2_i \geq7$, correspond to outliers in the galaxy properties. In other words, whether the outliers represent galaxies whose properties are on the tails (or outside) of the distribution. For this, we first consider all the galaxies in the Astrid training set. Next, for each galaxy, we take its stellar mass, gas mass, maximum circular velocity, and stellar metallicity (the four most important properties according to Fig. \ref{fig:ranking}). We then project these properties into 2D maps and show their distribution in Figs. \ref{fig:properties_Illustris}, \ref{fig:properties_SIMBA}, \ref{fig:properties_astrid}, and \ref{fig:properties_magneticum} with hexagons. The color of the hexagons indicates how many galaxies are in that region, as indicated in the color bar.

We then test the model, trained on Astrid galaxies, on galaxies from the IllustrisTNG, SIMBA, Astrid, and Magneticum test sets. For each galaxy, we compute the value of the $\chi^2_i$. We then select the ten galaxies with the highest $\chi^2$ value for each suite. Finally, we project the properties of these galaxies into the 2D plots. We show the results in Figs. \ref{fig:properties_Illustris}, \ref{fig:properties_SIMBA}, \ref{fig:properties_astrid}, and \ref{fig:properties_magneticum} with colored points. In the case of IllustrisTNG, SIMBA, and Magneticum galaxies, it is clear that those galaxies are far away from the main distribution of Astrid galaxies (sometimes mostly along one or two directions). In the case of Astrid, the galaxies are instead located within the distribution, with some of them touching the tails. This is reflected in their smaller $\chi^2$ values (see captions of Figs. \ref{fig:properties_Illustris}, \ref{fig:properties_SIMBA}, \ref{fig:properties_astrid}, and \ref{fig:properties_magneticum}).

To investigate whether the distribution of the galaxies with large $\chi^2_i$ values is different from the ones with smaller values, we have repeated the above exercise with galaxies randomly taken from the different test sets. We find that a larger fraction of the galaxies occupies regions in parameter space more densely covered by the Astrid galaxies. However, a random sampling of the IllustrisTNG, SIMBA, and Magneticum galaxies also selects galaxies that are located on the tails of the Astrid distribution. This is expected as the fraction of galaxies with $\chi^2_i\geq4$ can be 22\%, 38\%, and 43\% in the case of IllustrisTNG, SIMBA, and Magneticum, respectively.

While this is not a rigorous analysis, our results indicate that the model outliers represent galaxies whose properties are located in the tails of the distribution. This fact can be exploited to increase the robustness of the models. In future work, we plan to make use of more sophisticated machine learning tools, like normalizing flows, to address this question in a more rigorous manner.

\section{Robustness with two suites}
\label{sec:two_suites}

One of the reasons behind the lack of robustness of our models may be that galaxies from the different suites are just too different and their properties live in different regions. Training models on galaxies from two (or more) suites may improve the robustness of the model by forcing it to learn common features across models. On the other hand, training on such configuration may make the model to first classify the galaxy (e.g. recognizing it is a SIMBA galaxy) and then perform the usual task. If so, the model will not generalize well. 

\begin{figure*}[h!]
\centering
\includegraphics[width=1.0\linewidth]{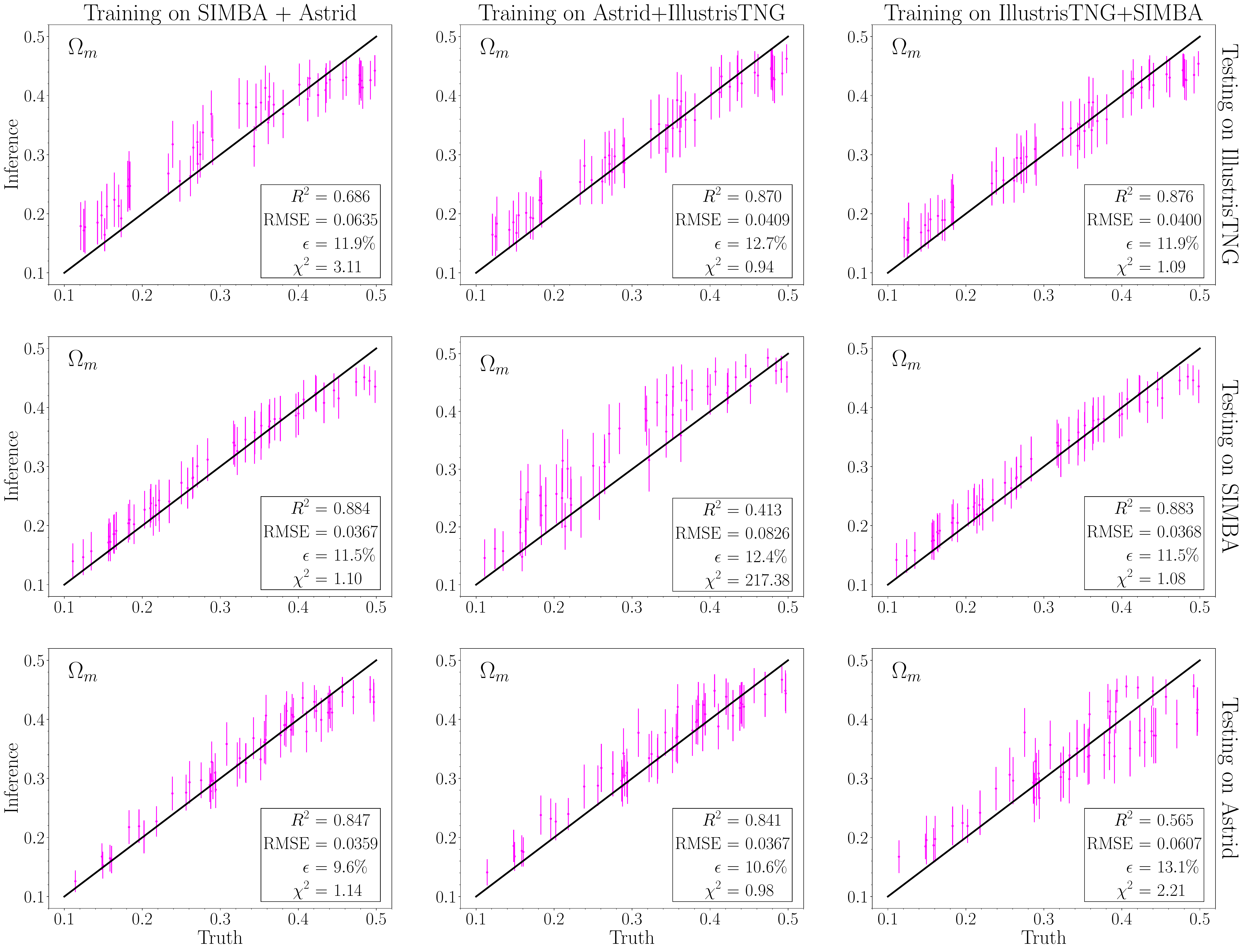}
\caption{Same as Fig. \ref{fig:robustness_no_chi2_cut} but training on SIMBA and Astrid galaxies (left column), Astrid and IllustrisTNG galaxies (middle column), and IllustrisTNG and SIMBA galaxies (right column). As can be seen, training on galaxies from two suites does not make the model robust.}
\label{fig:robustness_2suites}
\end{figure*}

In order to test this we have trained models using galaxies from two suites, e.g. Illustris and SIMBA galaxies. We then test the model on galaxies from all different suites. We show the results in Fig. \ref{fig:robustness_2suites}. We find that models trained on galaxies from two suites work well when tested on galaxies from those suites, but fail when tested on a third suite. This indicates that we cannot build robust models by training networks on galaxies from different simulations. Given the fact that the number of different subgrid physics models is really small, we believe this statement may hold in general.

\bibliography{references}{}
\bibliographystyle{aasjournal}

\end{document}